\documentclass[10pt,journal,compsoc]{IEEEtran} 
\usepackage{CJK}
\usepackage{amssymb}
\usepackage{amsmath}
\usepackage{graphicx, subfigure}
\usepackage{url, cite}
\usepackage{verbatim}
\usepackage{booktabs}
\usepackage{multirow}
\usepackage{bigstrut}
\usepackage{booktabs}
\usepackage{amsmath,cite,graphicx,times,subfigure,url,verbatim}
\usepackage{algorithmic}
\usepackage{algorithm}
\usepackage{epstopdf}

\newtheorem{subsec:coding}{subsec:coding}

\newcommand{\tabincell}[2]{\begin{tabular}{@{}#1@{}}#2\end{tabular}}

\begin{document}

\title{Cognitive-LPWAN: Towards Intelligent Wireless Services in Hybrid Low Power Wide Area Networks}

\author{Min~Chen,~Yiming~Miao,~Xin~Jian,~Xiaofei~Wang,~Iztok~Humar

\thanks{M. Chen and Y. Miao are with the School of Computer Science and Technology, Huazhong University of Science and Technology, Wuhan, China (minchen2012@hust.edu.cn, yimingmiao@hust.edu.cn).}
\thanks{X. Jian is with the College of Communication Engineering, Chongqing University, Chongqing, 400030, China (jianxin@cqu.edu.cn).}
\thanks{X. Wang is with the Tianjin Key Laboratory of Advanced Networking, College of Intelligence and Computing, Tianjin University, Tianjin 300350, China. (xiaofeiwang@tju.edu.cn).}
\thanks{I. Humar is with University of Ljubljana, Ljubljana, Slovenia (iztok.humar@fe.uni-lj.si).}
\thanks{Xiaofei Wang is the corresponding author.}
}

\markboth{Under review: IEEE Transactions on Green Communication and Networking, VOL. XX, NO. YY, MONTH 20XX}{}

\maketitle


\begin{abstract}
The relentless development of the Internet of Things (IoT) communication technologies and the gradual maturity of Artificial Intelligence (AI) have led to a powerful cognitive computing ability. Users can now access efficient and convenient smart services in smart-city, green-IoT and heterogeneous networks. AI has been applied in various areas, including the intelligent household, advanced health-care, automatic driving and emotional interactions. This paper focuses on current wireless-communication technologies, including cellular-communication technologies (4G, 5G), low-power wide-area (LPWA) technologies with an unlicensed spectrum (LoRa, SigFox), and other LPWA technologies supported by 3GPP working with an authorized spectrum (EC-GSM, LTE-M, NB-IoT). We put forward a cognitive low-power wide-area-network (Cognitive-LPWAN) architecture to safeguard stable and efficient communications in a heterogeneous IoT. To ensure that the user can employ the AI efficiently and conveniently, we realize a variety of LPWA technologies to safeguard the network layer. In addition, to balance the demand for heterogeneous IoT devices with the communication delay and energy consumption, we put forward the AI-enabled LPWA hybrid method, starting from the perspective of traffic control. The AI algorithm provides the smart control of wireless-communication technology, intelligent applications and services for the choice of different wireless-communication technologies. As an example, we consider the AIWAC emotion interaction system, build the Cognitive-LPWAN and test the proposed AI-enabled LPWA hybrid method. The experimental results show that our scheme can meet the demands of communication-delay applications. Cognitive-LPWAN selects appropriate communication technologies to achieve a better interaction experience.
\end{abstract}


\section{Introduction}

The Internet of Things (IoT) forms connections between people and things, as well as between things and things based on wired or wireless communication technologies. The IoT establishes a thing-thing Internet with a wide geographical distribution and provides innovative applications and services~\cite{K2017}. At present, global telecommunication operators have established mobile cellular networks with wide coverage~\cite{Cellular2017Asadi}. Although the 2G, 3G, 4G and other mobile cellular technologies~\cite{Cellular2017Vaze} support wide coverage and high transmission rates, they suffer from various disadvantages, such as large power consumption and high costs. According to Huawei's analysis report released in February 2016, the number of connections between the things and things on a global mobile cellular network occupies only 10\% of all the connections~\cite{Huawei}.

The crucial design purpose of mobile cellular-communications technology is to improve the interpersonal communication efficiency, since the current capacity of the mobile cellular network is not sufficient to support the massive connections between things and things. Up to 2025 the number of connections with industrial wireless sensing, tracking and control devices will reach to 500 million~\cite{K2017}. Therefore, the IoT should be able to provide end-users with convenient and efficient intelligent services and open access to historical data, while integrating a large number of heterogeneous terminal devices (the terminals of things) in a transparent and seamless manner. With such services, artificial intelligence (AI) systems can monitor the users and their surroundings more efficiently~\cite{brain}. This results in smart cities, smart homes, autopilot system~\cite{Vehicles2016Yu} and health-monitoring applications, and yields a greener, more environmentally friendly, and more efficient IoT ecosystem for more cost-effective systems.

\begin{figure*}
  \centering
  \includegraphics[width=5.2in]{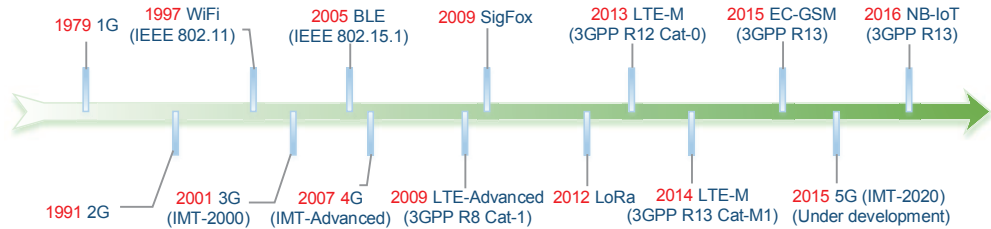}\\
  \caption{Historical evolution of wireless-communication technologies.}\ \label{fig1}
\end{figure*}

The diversity in terms of requirements and technologies for the Internet of Things has given rise to the heterogeneity of the network structures and the instability of the design solutions. Although traditional cellular networks provide long-distance coverage, they hardly provide energy-efficient (EE) connectivity due to their complex modulation and multiple access technologies~\cite{Cellular2016Ge}. As a result of the continuous development of the IoT its communication technologies have become more mature. These technologies can be divided into two main categories according to their transmission distances, as follows. The first category includes short-range communications technologies with representative technologies of Zigbee, WiFi, Bluetooth, Z-wave, etc. and typical application scenarios such as smart homes. The second category is the wide-area network communication technology defined as a low-power wide-area network (LPWAN), with typical application scenarios such as autonomous driving. The wide-area-network communication technology is required in low-speed businesses as a revolutionary IoT access technology, where typical wireless local area networks (WLANs), i.e., the WiFi, are less costly but suffer from a limited coverage distance. Therefore, LPWA technology is very promising thanks to its long-distance coverage, low power consumption, low data rates and low costs~\cite{lpwa}.

LPWA technologies can be classified into two categories. The first category includes the LoRa, SigFox and other technologies that work with unlicensed spectra. The second category includes 2/3/4G cellular communication technologies with licensed spectra based on the 3GPP, such as EC-GSM, LTE-M, NB-IoT, etc. According to such diverse wireless-communication technologies, here we propose the cognitive low-power wide-area network (Cognitive-LPWAN). Such a network aims at smart cities, green IoT and other heterogeneous networks in AI applications such as smart home, health monitoring, automatic driving and emotional interaction. We show that the Cognitive-LPWAN provides users with more efficient and convenient intelligent services by taking advantage of LPWA technologies.

Generally speaking, the purpose of a LPWAN is to provide long-distance communication, i.e., about 30 kilometers coverage in rural areas and 5 kilometers coverage in urban areas. Meanwhile, many IoT devices are characterized by a service time  of more than 10 years. Therefore, we need to improve the transmission range and power consumption of the LPWAN in order to adapt it to IoT applications that are highly extensible, e.g., the intelligent monitoring infrastructure, in which only a small portion of the data is transmitted. Thus, two possible technologies are proposed to solve these two problems. The first one is an ultra-narrowband technology that enhances the signal-to-noise ratio (SNR) by focalizing the signals in a narrow band. The narrowband IoT (NB-IoT)~\cite{nb1} is an implementation example of this approach. Another approach is to use a coding gain to alleviate the high noise and power consumption in wideband receivers, such as the Long-Range Wireless Communication (LoRa)~\cite{lora} technology that increases the transmission distance by increasing the power consumption. However, wireless-communication technologies with unlicensed spectra can conflict with other business flows with respect to channel collisions and spectrum occupation if they are not well controlled~\cite{Outage}. If this technology is abandoned for these reasons, a market with hundreds of millions of IoT terminals will also be lost.

Artificial intelligence technologies have become more mature, and we have now more powerful cognitive computing capabilities in regards to business perception at the user level, intelligent transmission at the network level and big-data analysis in the cloud. Aiming at traffic control, this paper proposes a new, intelligent solution for the Cognitive-LPWAN architecture, i.e., the AI-enabled LPWA hybrid method. We use the AI algorithm in a data and resource cognitive engine. LPWA technology has been widely applied to green IoT, so hopefully the new LPWAN architecture and its solutions for wireless-technology selection will contribute to the IoT ecosystem.

The main contributions of this paper are as follows:

\begin{itemize}
  \item We investigate several typical LPWA technologies, including the LoRa, SigFox and other wireless-communication technologies with unlicensed spectra, and 3GPP-supported 2/3/4G cellular-communication technologies with licensed spectra such as EC-GSM, LTE Cat-m, NB-IoT, etc.
  \item We propose the Cognitive-LPWAN as a combination of multiple LPWA technologies, ensuring more efficient and convenient user experiences in the AI services on the network layer. As a result, stable and efficient communication is guaranteed between the people and things, people and people, and things and things in the heterogeneous IoT.
  \item We puts forward the AI-enabled LPWA hybrid method, starting off from the perspective of the flow control, which makes a balance between the communication time delay and energy consumption in heterogeneous IoT devices. By using the AI algorithm, we achieve smart control for communication traffic, intelligent applications and services for the choice of the wireless-communication technology.
  \item We establish an experimental platform according to the AIWAC emotion interaction system, and compare the AI-enabled LPWA hybrid method with the single-technology mode. The experimental results show that our method can choose the communication technology accordingly and demonstrates proper transmission-delay performance.
\end{itemize}

The rest of this paper is organized as follows. Section~\ref{sec:tech} summarizes the existing heterogeneous, low-power, wide-area-network technologies. Section~\ref{sec:arch} introduces Cognitive-LPWAN and proposes the solution of multiple wireless-communication technologies in the smart-cities environment. Next, we present the AI-enabled LPWA hybrid method modeling in Sec.~\ref{sec:method}. Section~\ref{sec:testbed} demonstrates the building testbed through different wireless-communication technologies based on baby robots. This building testbed is used to test the performance of the proposed AI-enabled LPWA hybrid method. We then discuss open issues for the future in Sec.~\ref{sec:open}. Finally, Sec.~\ref{sec:conclusion} summarizes the paper.

\section{Heterogeneous Low-Power Wide-Area-Network Technology}\label{sec:tech}

\subsection{SigFox}

SigFox is provided by the SIGFOX company that was founded by the Ludovic Le Moan and Christophe Fourtet in 2009. As a LPWA technology worked on non-licensed spectrum, SigFox has been rapidly commercialized and provides network devices with ultra-narrowband technology~\cite{sigfox}. Different from mobile communication technology, it is a protocol that aims to create wireless IoT special networks with low power consumption and low costs. More concretely, the maximum length of each message on a device that operates based on the Sigfox network is nearly 12 bytes due to the 50 microwatts upper limit of one-way communications. And no more than 150 messages will be sent per day by each device. Moreover, the coverage provided by a SigFox network can reach up to 13 kilometers.

\subsection{LoRa}

LoRa is currently one of the most common LPWA technologies worked on non-licensed spectrum, which is provided by SemTech~\cite{lora}. The main characteristics of LoRa wireless technology include 20 km coverage range, 10,000 or even millions of nodes, 10 years of battery life, and 50 kbps data rate. As a wireless technology based on the sub-GHz frequency band, LoRa Technology enables massive smart IoT applications and solve some of the biggest challenges facing large-scale IoT, including energy management, natural resource reduction, pollution control, infrastructure efficiency, disaster prevention, and more. With an increase in the number of LoRa devices and networks and consequent mutual spectrum interferences, a unified coordination-management mechanism and a large network are required to allocate and manage the communication spectra due to the unauthorized spectrum of the LoRa. Some aspects should be considered in LoRa applications, such as the transmission distance, the number of connected nodes, application scenarios, the power consumption and costs.

\begin{figure}
  \centering
  \includegraphics[width=3.5in]{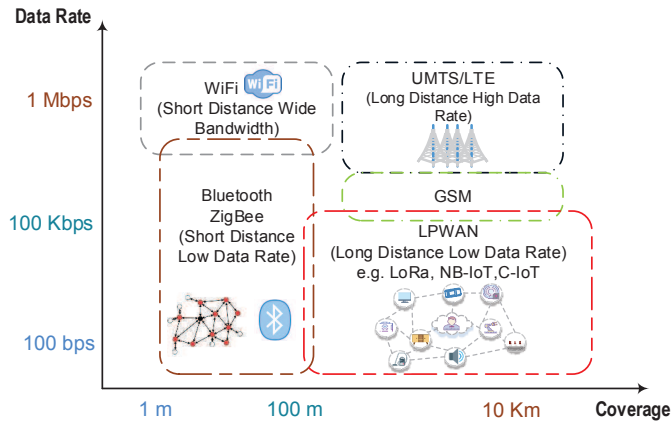}\\
  \caption{Heterogeneous wireless-communication technologies.}\ \label{fig2}
\end{figure}

\subsection{LTE-M}

3GPP possesses three basic standards, i.e., LTE-M~\cite{ltem}, EC-GSM~\cite{EC-GSM} and NB-IoT~\cite{nb2}, which are, respectively, based on the LTE evolution, GSM evolution and Clean Slate technologies. Start from 3GPP R12 in 2013, LTE-Machine-to-Machine (LTE-M) is intended to meet the requirements of IoT devices based on the existing LTE carriers with an upstream and downstream data rate of 1 Mbps~\cite{ltem}. LTE-M possesses four basic advantages of LPWA technology including wide coverage, massive connections, low power consumption and low module costs. Due to the wide coverage, LTE-M achieves a transmission gain of 15 dB in comparison to the existing technologies under the same licensed spectrum (700 - 900 MHz), which improves the coverage ability of an LTE network significantly. Besides, a LTE-M network cell can support nearly 100,000 connections. The standby time for LTE-M terminals can be up to 10 years.

\subsection{EC-GSM}

EC-GSM (Extended Coverage-GSM)~\cite{EC-GSM} was put forward after the narrowband IoT technology transferring to GSM (Global System for Mobile Communication), according to the research project in the 3GPP GERAN (GSM EDGE Radio Access Network) in 2014. As a result, a wider coverage was achieved with 20 dB higher than traditional GPRS, and five major objectives were proposed as follows: the improvement of the indoor coverage performance, support to large-scale device connectivity, simplification of the devices, reduction of the power consumption and latency. However, with the continuous development of the technology, the CIoT (cellular IoT) communication should be redefined, resulting in the emergence of NB-IoT. NB-IoT is a clean-slate solution which is not based on GSM. Therefore, the research of the cIoT was transferred to the RAN group, that will be introduced in next subsection. The EC-GSM is continued to be developed by GERAN until the 3GPP R13 NB-IoT standard is almost updated.

\begin{table*}
  \centering
  \caption{Comparison of LPWA technologies}
  \begin{tabular}{|c|c|c|c|c|c|c|}
    \hline
        \multirow{3}{*}{Technologies}&LTE-Evolution&\multicolumn{3}{|c|}{Narrowband}&\multicolumn{2}{|c|}{Non-3GPP}\\
        \cline{2-7}
            &\multirow{2}{*}{LTE-M}&\multicolumn{2}{|c|}{NB-IoT}&\multirow{2}{*}{EC-GSM}&\multirow{2}{*}{LoRa}&\multirow{2}{*}{SigFox}\\
        \cline{3-4}
            & &NB-LTE&NB-CIoT& & & \\
    \hline
        Coverage&$<$ 11 km&$<$ 15 km&$<$ 15 km&$<$ 15 km&$<$ 20 km&$<$ 13 km\\
    \hline
        Spectrum&\tabincell{c}{Licensed \\(7-900 MHz)}&\tabincell{c}{Licensed \\(7-900 MHz)}&\tabincell{c}{Licensed \\(8-900 MHz)}&\tabincell{c}{Licensed \\(7-900 MHz)}&\tabincell{c}{Unlicensed \\(867-869 MHz or \\ 902-928 MHz)}&\tabincell{c}{Unlicensed \\(900 MHz)}\\
    \hline
        Bandwidth&1.4 MHz&200 kHz&200 kHz&2.4 MHz&125 kHz, 250 kHz, 500 kHz&100 kHz\\
    \hline
        Date Rate&$<$ 1 Mbps&$<$ 150 kbps&$<$ 400 kbps&10 kbps&$<$ 50 kbps&$<$ 100 bps\\
    \hline
        Battery Life&$>$ 10 years&$>$ 10 years&$<$ 10 years&$>$ 10 years&$<$ 10 years&$>$ 10 years\\
    \hline
  \end {tabular}
  \label{tab1}
\end{table*}

\subsection{NB-IoT}

In 2015, the 3GPP RAN initiated research on a new air-interface technology called Clean Slate CIoT for narrowband wireless access, which covered the NB-CIoT and NB-LTE. These two technologies are fairly incompatible and compatible with the existing LTE for easier deployment, respectively. In 2016, for a unified solution, NB-IoT was considered as a fusion of NB-CIoT and NB-LTE by 3GPP R13. NB-IoT is a new type of LPWA technology intended for sensing and data collection applications, such as intelligent electric meters, environment supervision, and etc. It can satisfy the requirements of non-latency-sensitive and low-bitrate applications (time delay of uplink can be extended to more than 10 s, and uplink or downlink for a single user are supported at 160 bit/s at least), which are coverage enhancement (coverage capacity is increased 20 dB), ultralow power consumption (a 5-Wh batter can be used by one terminal for 10 years), and massive terminal access (a single sector can supports 50,000 connections) at transmission bandwidth of 200 kHz.

\subsection{Comparison}

We summarize the development time line of the wireless communication technology. Fig.~\ref{fig1} shows the evolution process of 1G, 2G, 3G, BLE, 4G, SigFox, LoRa, LTE-M, EC-GSM, 5G, NB-IoT, etc. At the same time, we also have the joint-development nodes of other technologies, such as the mobile and short-distance communication technologies. It is clear that, at present, wireless-communication technology is developing vigorously, and represents a key node to deploy and use the hybrid LPWAN architecture.

According to the IoT standards of 3GPP~\cite{r14,3GPPRAN} and technology development presentation of Dr. Wei~\cite{wei2016}, we summarizes and compares the mentioned LPWA technologies in terms of the coverage, spectrum, bandwidth, data rate, and battery life in Table~\ref{tab1}. LTE possesses the smallest coverage because of its large data transmission rate and high bandwidth resources and energy consumption. Hence, it is not suitable for WAN applications. LoRa does not exhibit a fixed-frequency spectrum and bandwidth resources, because it works in the unauthorized frequency band, and is influenced by the regulations in the area they are used in. As a result, it is not suitable for mobile communications, whereas it is suitable for wireless sensor network communications and private businesses. A low power consumption is the primary requirement for all LPWA technologies, so the above technologies provide a battery life of up to 10 years.

Fig.~\ref{fig2} compares the coverage and the data rate of the commonly used LPWA technology and other wireless-communication technologies, e.g., LoRa, EC-GSM, NB-IoT, WiFi, BLE (bluetooth low-power consumption), and LTE-M. Short-distance and high-bandwidth communication technologies, such as WiFi, can cover up to 100 m with a data transmission rate of 100 Mbps. This communication mode is suitable for short-distance and high-bandwidth applications. For short-distance and low-data-transmission-rate communication technologies, e.g., Bluetooth and ZigBee, the highest coverage can reach to 100 m or so while its data transfer rate is 100 kbps. The communication mode is suitable for short-distance, low-bandwidth applications. Long-distance and high-data-transmission-rate communication technologies, such as UMTS and LTE, can cover a maximum range of 10 km, with a data transmission rate of 100 Mbps. This communication method is suitable for long-distance and high-bandwidth applications. GSM can provide coverage up to 10 km, and a data-transmission rate of close to 100 kbps. This communication mode is suitable for long-distance and medium-bandwidth applications. Long-distance low-data-transmission-rate communication technologies, such as LoRa, NB-IoT, C-IoT and NB-CIoT, can cover up to 10 km, while the data transmission rate is 100 kbps, and this communication method is suitable for long-distance and low-bandwidth applications.

\section{Cognitive Low-Power Wide-Area-Network Architecture}\label{sec:arch}

\begin{figure*}
  \centering
  \includegraphics[width=6.5in]{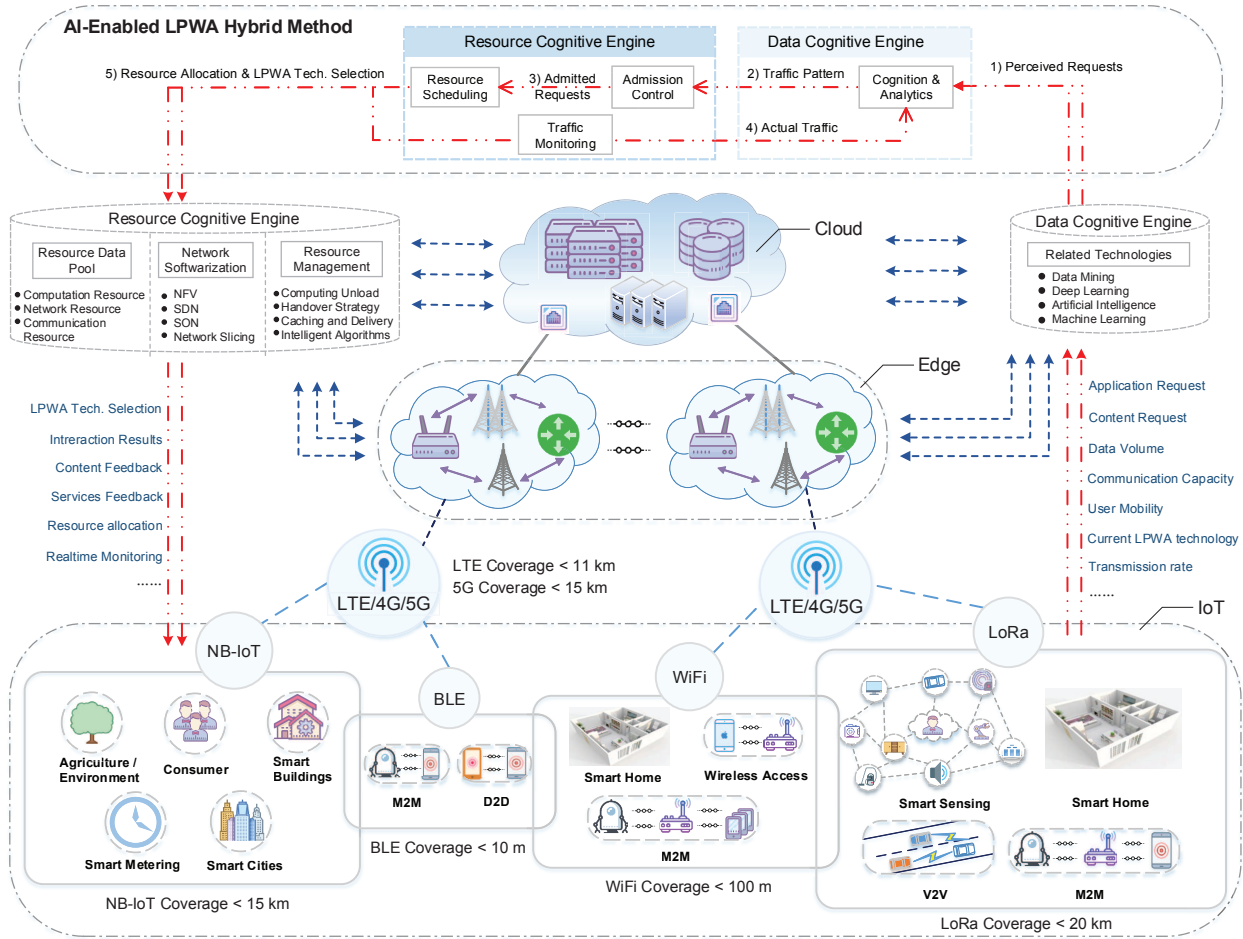}\\
  \caption{Architecture of the cognitive low-power wide-area network.}\ \label{fig3}
\end{figure*}

Fig.~\ref{fig3} shows the architecture of Cognitive-LPWAN. This architecture takes advantage of the LPWA communication technology (in Section~\ref{sec:tech}), the heterogeneous IoT applications, SDN and AI technology. Cognitive-LPWAN realizes the mixing of a variety of LPWA technologies, and provides the users with more efficient and convenient intelligent services. The main applications of this technology are smart cities, green IoT, general heterogeneous networks, as well as AI applications such as smart home, health monitoring, automatic driving and emotional interaction.

\textbf{(1) IoT / Heterogeneous LPWANs}

It is clear from the above text that, at present, there are many types of wireless-communication technologies causing heterogeneous and complicated IoT infrastructure and applications. As shown in Fig.~\ref{fig3}, the heterogeneous IoT platform based on a variety of wireless-communication technologies, including the LPWA technology. These wireless-communication technologies include unlicensed LPWA technology (LoRa, SigFox), short-distance wireless-communication technology (BLE, WiFi), and mobile cellular-communications technology (NB-IoT, LTE, 4G and 5G). The applications and coverage areas supported by these technologies intersect and are widely used in users' lives. Technically, they can even replace each other. However, for a comprehensive consideration of cost and communication performance, power consumption, mobility~\cite{Mobility2018Mancuso} and other factors, the applications of these techniques still have their own advantages in specific scenarios. The coverages of these technologies, listed in Fig.~\ref{fig3}, are as follows. The coverage of NB-IoT is $<$ 15 km, the coverage of BLE is $<$ 10 m, the coverage of WiFi is $<$ 100 m, the coverage of LoRa is $<$ 20 km, the coverage of LTE is $<$ 11 km and the coverage of 5G is $<$ 15 km~\cite{5GMobile}. The above coverage areas (as well as other performance indicators, such as transmission rate, bandwidth, sensor capacity are discussed in Sec.~\ref{sec:tech}) limit the popularity of these technologies throughout the field. Nevertheless, different applications or services have different requirements and use different communication technologies. For instance, BLE is often used for short-distance mobile-phone or robot communications, and WiFi is low cost, has a fast transmission rate and is a stable communication in WLAN (wireless WAN network) applications such as the smart home. Moreover, Nb-IoT base-stations can be built based on LTE and 4G infrastructures, and use LTE spectrum resources, which greatly saves the promotion costs and supports low-power-consumption WAN applications. LoRa is widely used for smart sensing (the interconnection and transmission of mass sensors) and other applications thanks to its advantages in working in a non-authorized spectrum. The IoT and its intelligent applications are given in details in Fig.~\ref{fig3}. Among them, NB-IoT supports agriculture and environment monitoring, consumer tracking, smart buildings, smart metering and smart cities. BLE supports machine-to-machine (M2M) communication, device-to-device (D2D) communication and other applications. WiFi supports smart home, wireless access~\cite{wireless2015Vaze} and M2M communication and other applications; LoRa supports smart sensing, smart home, vehicle-to-vehicle (V2V) communication and M2M communication. The LTE, 4G and 5G, as a wireless-access technology on the Edge of the IoT, connect the above IoT cells (NB-IoT, BLE, WiFi, LoRa, etc.) that are the closest to the user to the Edge Cloud and realize the network-access function.

\textbf{(2) Cognitive engine}

It is clear from Fig.~\ref{fig3} that in the Edge Cloud and Cloud~\cite{cloud2017guo} we have introduced the cognitive engine, deployed high-performance artificial intelligence algorithms, and stored a lot of user data and IoT business flows. Consequently, it can provide a high-precision calculation and data analysis and provide Cloud support for the selection of the LPWA communication technology. The Cognitive engine is divided into two types, i.e., the resource-cognitive engine and the data-cognitive engine~\cite{M2018}. 1) Data cognitive engine: processing the real-time multi-modal business data flow in the network environment, with data analysis and business automatic processing power, executing the business logic intelligently, and realizing the cognition to the business data and resource data through a variety of cognitive calculating methods. This includes data mining, machine learning, deep learning and artificial intelligence. As a result, a dynamic guide~\cite{Dynamic2018Tulino} of the resource-allocation~\cite{resource2017shan,Allocation2016Tulino} and cognitive services will be achieved. 2) The resource-cognitive engine can perceive the computing resources, communication resources~\cite{Genetic2015shan} and network resources~\cite{Dynamic2016Ji} of heterogeneous resources IoT~\cite{Resource2016Ji}, edge cloud and remote cloud~\cite{Virtualization2018Yu}, and make the real-time feedback of the comprehensive resources data to the data cognitive engine. Meanwhile, the network resources include the network type, data flow, communication quality of business and other dynamic environment parameters. Added to this, the analysis results of the data-cognition engine are received to guide the selection of the LPWA technologies and real-time dynamic optimization and allocation of resources.

\textbf{(3) AI-enabled LPWA hybrid method}

The AI-LPWA hybrid method is a key component of the new architecture proposed in this paper, as shown in the circular flow chart in Fig.~\ref{fig3}. In this section we discuss the algorithm flow. When a user or device in the IoT makes a request, the business flow is transferred to the edge of the IoT through the current LPWA technology. It is then transmitted by the LTE, 4G, 5G~\cite{5G2016Ge} and other technologies to the Edge Cloud (wireless access point, router, base-station and other Edge computing nodes). If its request computation amount exceeds the Edge Cloud's capability, it will be forwarded to the Cloud~\cite{DelayA2017Ji} again by the Edge computing node. In addition, the question is whether the Edge Cloud or the Cloud dealing with the business flow, the data-cognitive engine we deployed in each compute node will perceive all the information contained in the business flow, and consolidate the perceived requests, including the application request, content request, data volume, communication capacity, user mobility, current LPWA technology, and transmission rate. Subsequently, the data-cognitive engine will intelligently analyze the perceived requests, and extract the traffic pattern to transmit it to the resource-cognitive engine for the admission control. When the Edge Cloud computing node/Cloud computing node admits the requests, the cognitive engine will allocate the computing resources for the IoT users or devices launching the request, and determine what type of LPWA wireless-communication technology will be adopted to carry the information back. The returning information (business content and control information) includes the LPWA technology selection, interaction result, content feedback, services feedback, resource allocation, and real-time monitoring. Among them, the selected LPWA technology will be used in the next request until the feedback traffic pattern of the reverse propagation in the next traffic monitoring does not conform to the current technological performance. At this point it will be replaced. The specific AI-enabled LPWA hybrid method modeling will be introduced in Section~\ref{sec:method}.

\section{AI-Enabled LPWA Hybrid Method Modeling}\label{sec:method}

This section presents the mathematical model of the AI-LPWA hybrid method. Here, we introduce the no-label learning algorithm for modeling~\cite{LabelLess}. Specifically, in this paper, aiming at the current business flow reaching the compute nodes, its traffic patterns are extracted and added to the traffic-pattern data set already there in the data-cognitive engine. Next, we integrate the existing data to predict and evaluate the selection of wireless-communication technologies in this business. We assume that the existing traffic-pattern data set (i.e., the labeled data sets) are $x^l=(x_1^l,x_2^l,...,x_n^l)$, where $x$ represents the traffic pattern (including the communication time delay, data volume, transmission rate, user mobility, and computing complexity), $n$ represents the number of labeled data. The service request (business flow) data set (i.e., unlabeled data set) that reaches the compute node is $x^u=(x_1^u,x_2^u,...,x_m^u)$, where $m$ represents the number of non-labeled data. The label corresponding to the label data set is $y^l=(y_{x_1}^l,y_{x_2}^l,...,y_{x_n}^l)$. Suppose the probability of assigning some wireless communication technology to this traffic flow is $y_{x_i}^u=\{p_{x_i^u}^1,p_{x_i^u}^2,...,p_{x_i^u}^c\}$. Here, $p_{x_i^u}^j$ represents the probability that the traffic pattern $x_i^u$ is predicted as category $j$, $j$ represents the wireless-communication technology, where $j=\{LTE, NB-IoT, LoRa, Sigfox, BLE, WiFI, 4G, 5G...\}$, $c$ represents the number of $j$. Then, the pre-selection probability entropy of the communication technology of the non-labeled data set is shown as Eq.~(\ref{eq1}).

\begin{equation}\label{eq1}
E(y_{x_i}^u)=-\sum_{j=1}^c{p_{x_i^u}^j{log(p_{x_i^u}^j)}}
\end{equation}

It is clear from the above formula that decreasing the entropy value results in a lower prediction uncertainty for the new annotated data (i.e., the wireless-communication technology used to allocate the business flow). Therefore, entropy can be used as the pre-selection standard of the LPWA technology. However, the selection of the threshold value is uncertain, i.e., the low entropy value might still lead to the wrong selection of the communication technology (which may be directly reflected by the user experience data, such as interactive delay and energy consumption). If the pre-selection of the LPWA technology is not accurate, i.e., this traffic pattern is added to the existing data set as trusted data, this can lead to the accumulation of errors in the subsequent training model. Specifically, the label corresponding to the traffic-pattern data was incorrectly marked, which results in increased noise added to the data set and increases the forward propagation error.

In order to overcome this problem, we apply the traffic monitoring to the business flow, i.e., each time there is a new data joining the training set, the data joining each time will be evaluated. However, we might not always trust low-threshold data. Assume the independent tag data set added based on the low-entropy threshold is $z$. Then, for any $x_i^u\subseteq{z}$, the following conditions should be met, as shown in Eq.~(\ref{eq2}).

\begin{equation}\label{eq2}
E(y_{x_i}^u)\leqslant{E(y_{x_i}^l)}
\end{equation}

In other words, the data of the error markers can be corrected through a re-evaluation, i.e., reverse propagation in Fig.~\ref{fig3} (4), and the tag results might be fine-tuned to reduce the error.

\section{Testbed and Analysis}\label{sec:testbed}

\begin{figure}
  \centering
  \includegraphics[width=3.5in]{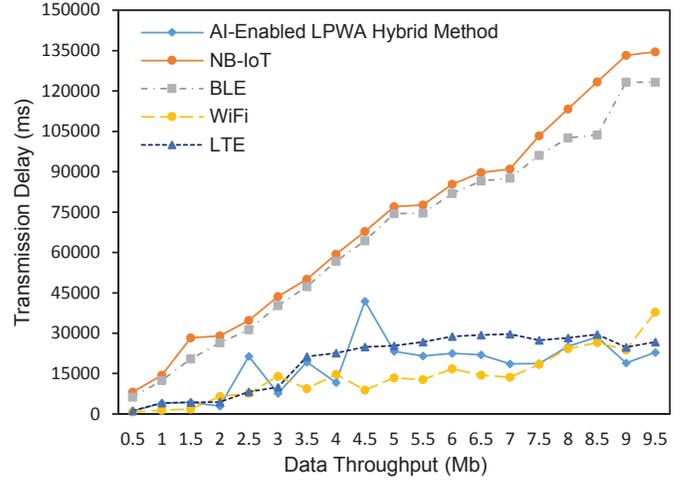}\\
  \caption{Comparison of the transmission delay between the AI-enabled LPWA hybrid method and different wireless-communication technologies and in the AIWAC system.}\ \label{fig4}
\end{figure}

\begin{figure}
  \centering
  \includegraphics[width=3.5in]{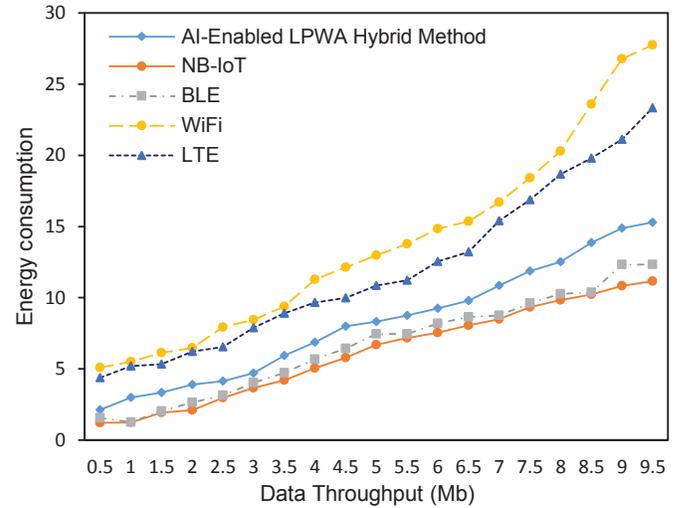}\\
  \caption{Comparison of the energy consumption between the AI-enabled LPWA hybrid method and different wireless-communication technologies and in the AIWAC system.}\ \label{fig5}
\end{figure}

We use the AIWAC system~\cite{aiwac} as an AI-enabled, heterogeneous, low-power, wide-area-network architecture (application scene), which realizes the emotional interaction between the users and the smartphone / AIWAC robot. In our real experimental environment, smart clients (smartphones, robots, etc.) represent the IoT devices, and the local server represents the deployed edge computing node. Meanwhile, the cloud deploys a big-data center and an analytics server as the remote service background of the smart client. Some of the hardware parameters include the Smart Client (Android 7.0, 8-core, 1.2 GHz frequency), Local Server (CentOS 7, Quad-core, 3.4 GHz frequency), GPU Analytics Server (Ubuntu 16.04, NVIDIA GTX1080ti*2) and Big data Center (Android 4.4, 16core 4 GHz + 42core 4 GHz). The computation capability of the Smart Client (Edge Computing Node) $<$ Local Server (Fog Computing Node) $<$ Big Data Center (Cloud).

We deployed several wireless-communication technologies, which include NB-IoT, BLE, WiFi and LTE, as mentioned in Section IV on smartphones (communication protocol alternatives) and AIWAC robots (hardware circuit-board communication modules). These technologies represent the current typical LPWA communication protocols and provide us with a performance basis for the communication protocols in a real environment for the proposed Cognitive-LPWAN. In addition, we deployed the AI-enabled LPWA hybrid method on the cognitive engine of the Smart Client, Edge Computing Node and Cloud Computing Node to test the interaction performance comparison between the proposed scheme and the one under the single communication mode.

Fig.~\ref{fig4} and Fig.~\ref{fig5} show the experimental results of the transmission delay and energy consumption with the data-transmission amount (network throughput) of different wireless communication technologies. Due to the narrow bandwidth of the NB-IoT and BLE technologies, their low transmission rate leads to a large interaction delay. In particular, they are seriously defective when transferring large content or files. However, they consume ultra-low energy consumption while transmitting small amounts of data, so they are often used in the low-power-consumption business. Data-transmission rates are an advantage for WiFi and LTE technologies, so transmission delays are less important than the two mentioned above. However,the energy consumption of LTE and WiFi are opposites. With the increase in energy consumption and costs, they represent a better option for businesses that are sensitive to the time delay. The transmission delay of the proposed AI-enabled LPWA hybrid method is under AIWAC testbed and is basically consistent with WiFi and LTE. This is due to the fact that the AIWAC system provides the emotion recognition and interactive services. And it is a time-delay-sensitive application. For the loss under the compromise between the energy consumption and delay of the choice, it is better to use a high-speed communication technology. For the compromise between the energy consumption and the time delay, the loss of the partial energy consumption is chosen as the cost, so it is preferential to be used in high speed communication technology.

\section{Open Issues}\label{sec:open}

Although Cognitive-LPWAN architecture is proposed in this paper, business shunt problems are discussed and the AI-enabled LPWA hybrid method was introduced, in the large-scale use of the hybrid LPWA technologies, the following challenging problems should be solved.

\textbf{(1) Security requirements: }There are many similarities and differences between the security requirements of Cognitive-LPWAN and conventional IoT. The security requirements of Cognitive-LPWAN mainly includes the hardware for the low-power IoT, the network communication mode, and the actual business requirements related to the equipment and other aspects. For instance, any tiny security breach could cause serious safety accidents in the IoT platform with a very large number of users~\cite{Secrecy,Privacy2017Guo}.

\textbf{(2) Energy consumption and real-time management: }Since LPWA technology has mostly been applied to the low-power long-distance service, its transmission rate is much lower than other technologies. In the heterogeneous IoT, considering the mobility of the users and the diversity of applications, we should compromise the energy consumption and real-time performance of multiple intelligent services. Accordingly, we introduce short-distance communication technologies, e.g., BLE and WiFi, as well as mobile cellular networks, e.g., 4G and 5G. We therefore meet the different requirements of various businesses for energy consumption and real time.

\textbf{(3) Spectrum resources optimization: }Since the LPWAN includes technologies working in an unauthorized spectrum (taking LoRa and SigFox, for example), limited spectrum resources are required to serve a large number of IoT devices. Hence, there are conflicts between the bandwidth occupation and the spectrum resources for a large number of communication users. Therefore, we need to use AI technologies such as machine learning to conduct concurrent control and schedule to similar users in the channel allocation, interference management, transmission power optimization and other aspects.

\textbf{(4) Infrastructure deployment: }The Cognitive-LPWAN deployment with distributive algorithms is a hot research topic at present. How to build a new LPWAN architecture on the layout of the existing infrastructure and achieve the goal of minimizing cost and maximizing utilization rate is the key point that telecom operators need to consider.

\section{Conclusion}\label{sec:conclusion}

The diversity of the Internet of Things (IoT) in terms of demand and technology has led to the heterogeneity of the network structure and the instability of the design scheme. This paper has compared the performance of the commonly used wireless-communication technologies. In particular, we considered different aspects, advantages and disadvantages of the LoRa, SigFox, LTE-M, EC-GSM and NB-IoT technologies which are typical in LPWA technology. Their advantages and disadvantages in terms of coverage, frequency spectrum, bandwidth and data rate, and battery life are discussed. In LPWA and short-distance communication technologies (BLE, WiFi) as well as mobile cellular-network technology (4G, 5G) are summarized. We aimed at wireless-communication technologies and heterogeneous networks, including smart cities and green IoT, as well as AI applications including smart home, health monitoring, automatic driving and emotional interaction. We next proposed Cognitive-LPWAN, which realizes the mixing of a variety of LPWA technologies and provides more efficient and convenient smart services. In addition, on the basis of the powerful cognitive computing (i.e., service awareness at the user level, intelligent transmission at the network level and a large data-analysis ability in the cloud) provided by the AI technology, we proposed the AI-enabled LPWA hybrid method from the perspective of traffic control. It uses the AI algorithms to conduct business shunt and filtering to wireless-communication technologies, intelligent applications and services for the choice of the wireless-communication technologies. Then, we took the AIWAC emotion interaction system as an example and built the architecture of Cognitive-LPWAN to test the proposed AI-enabled LPWA hybrid method. The experimental results show that our scheme can meet the demands of the communication (transmission) delay (and energy consumption) in applications where appropriate communication technologies are chosen to achieve a better interaction experience. Finally, we presented different aspects of future research regarding the security requirements, energy consumption and real-time management, spectrum resource optimization and infrastructure deployment.


\ifCLASSOPTIONcompsoc
\else
\fi


\bibliographystyle{IEEEtran}

\begin{thebibliography}{}


\bibitem{K2017}
K. Hwang, M. Chen, ``Big Data Analytics for Cloud/IoT and Cognitive Computing,'' \emph{Wiley}, U.K., ISBN: 9781119247029, 2017.

\bibitem{Cellular2017Asadi}
A. Asadi, V. Mancuso, R. Gupta, ``DORE: An Experimental Framework to Enable Outband D2D Relay in Cellular Networks'', \emph{IEEE/ACM Transactions on Networking}, No. 99, pp. 1--14, 2017.

\bibitem{Cellular2017Vaze}
R. Vaze, S. Iyer, ``Capacity of Cellular Wireless Network'', \emph{WIOPT 2017}, pp. 1--8, 2017. DOI:10.23919/WIOPT.2017.7959907.

\bibitem{Huawei}
Huawei, ``2016 is the Key Year for the Development of NB-IoT Industry,'' \emph{MWC2016}, 2016. URL: http://iot.ofweek.com/2016-02/ART-132209-8120-29069354.html

\bibitem{brain}
H. Lu, Y. Li, M. Chen, et al., ``Brain Intelligence: Go beyond Artificial Intelligence'', \emph{Mobile Networks \& Applications}, Vol. 23, No. 2, pp. 368-375, 2017.

\bibitem{Vehicles2016Yu}
F. R. Yu, ``Connected Vehicles for Intelligent Transportation Systems,'' \emph{IEEE Trans. Veh. Tech., Editorial}, Vol. 65, No. 6, pp. 3843--3844, June 2016.

\bibitem{Cellular2016Ge}
X. Ge, Y. Qiu, J. Chen, M. Huang, et. al, ``Wireless Fractal Cellular Networks,'' \emph{IEEE Wireless Communications}, Vol. 23, No. 5, pp. 110--119, Oct. 2016.

\bibitem{lpwa}
R. Usman, P. Kulkarni, and M. Sooriyabandara. ``Low Power Wide Area Networks: An Overview.'' \emph{IEEE Communications Surveys \& Tutorials}, Vol. 19, No. 2, pp. 855-873, 2017.

\bibitem{nb1}
M. Chen, Y. Miao, Y. Hao, K. Hwang, ``Narrow Band Internet of Things,'' \emph{IEEE Access}, Vol. 5, pp. 20557-20577, 2017.

\bibitem{lora}
M. Saari, A. M. Baharudin, P. Sillberg, S. Hyrynsalmi, W. Yan. ``LoRa-A survey of recent research trends,'' \emph{MIPRO 2018}, Opatija, Jadranska obala, Hrvatska, 21-25 May 2018.

\bibitem{Outage}
G. Wang, W. Xiang, J. Yuan, ``Outage Performance for Compute-and-Forward in Generalized Multi-Way Relay Channels,'' \emph{IEEE Communications Letters}, Vol.16, No.12, pp.2099-2102, Dec. 2012.

\bibitem{sigfox}
B. Vejlgaard, M. Lauridsen, H. Nguyen, I. Z. Kovacs, P. Mogensen; M. Sorensen, ``Coverage and Capacity Analysis of Sigfox, LoRa, GPRS, and NB-IoT,'' \emph{IEEE VTC Spring 2017}, Sydney, Australia, 4¨C7 June 2017.

\bibitem{ltem}
R. Ratasuk, et al. ``Narrowband LTE-M System for M2M Communication,'' \emph{IEEE VTC Spring 2014}, Seoul, South Korea, 18¨C21 May 2014.

\bibitem{EC-GSM}
S. Lippuner, B. Weber, M. Salomon, M. Korb, Q. Huang, ``EC-GSM-IoT network synchronization with support for large frequency offsets,'' \emph{IEEE WCNC 2018}, Barcelona, Spain, 15-18 April 2018.

\bibitem{nb2}
Y. Miao, W. Li, D. Tian, M. S. Hossain, M. F. Alhamid, ``Narrow Band Internet of Things: Simulation and Modelling,'' \emph{IEEE Internet of Things Journal}, Vol. 5, No. 4, pp. 2304-2314, 2017.

\bibitem{r14}
3GPP, ``Standards for the Iot'', 2 December 2016. URL: http://www.3gpp.org/news-events/3gpp-news/1805-iot\_r14

\bibitem{3GPPRAN}
P. Reininger, ``3Gpp standards for the Internet-of-Things,'' Huawei, Shenzhen, China, Tech. Rep. report no. 3GPP RAN WG 3, 2016.

\bibitem{wei2016}
J. Wei, ``Development status of 3GPP NB-IoT Internet of things technology,'' \emph{2016 Institute for Information Industry}, 2016.

\bibitem{Mobility2018Mancuso}
G. Rizzo, V. Mancuso, S. Ali, M. A. Marsan, ``Stop and Forward: Opportunistic Local Information Sharing Under Walking Mobility'', \emph{Ad Hoc Networks}, May 2018. DOI: 10.1016/j.adhoc.2018.05.011.

\bibitem{5GMobile}
W. Xiang, K. Zheng, X. Shen, ``5G Mobile Communications'', \emph{Springer}, 2017, ISBN: 978-3-319-34206-1.

\bibitem{wireless2015Vaze}
R. Vaze, ``Random Wireless Networks: An Information Theoretic Perspective,'' \emph{Cambridge University Press}, 2015.

\bibitem{cloud2017guo}
Q. Zhang, S. Guo, ``Online Shuffling with Task Duplication in the Cloud'', \emph{ZTE Communications}, Vol. 15, No. 4, pp. 38--42, October 2017.

\bibitem{M2018}
M. Chen, Y. Tian, G. Fortino, J. Zhang, I. Humar, ``Cognitive Internet of Vehicles,'' \emph{Computer Communications}, Vol. 120, pp. 58-70, May 2018.

\bibitem{Dynamic2018Tulino}
P. Hassanzadeh, A. M. Tulino, J. Llorca, E. Erkip, ``On Coding for Cache-Aided Delivery of Dynamic Correlated Content'', \emph{IEEE Journal on Selected Areas in Communications}, June 2018. DOI: 10.1109/JSAC.2018.2844579.

\bibitem{resource2017shan}
H. Shan, Y. Zhang, W. Zhuang, A. Huang, Z. Zhang, ``User Behavior-aware Scheduling based on Time-frequency Resource Conversion,'' \emph{IEEE Trans. Vehicular Technology}, Vol. 66, No. 9, pp. 8429--8444, Sept. 2017.

\bibitem{Allocation2016Tulino}
L. Jiao, A, M, Tulino, J. Llorca, A. Sala, `` Smoothed Online Resource Allocation in Multi-Tier Distributed Cloud Networks'', \emph{IEEE IPDPSW 2016}, Chicago, USA, 23-27 May 2016.

\bibitem{Genetic2015shan}
H. Shan, Z. Ye, Y. Bi, A. Huang, ``Genetic Algorithm based Resource Management for Cognitive Mesh Networks with Real-time and Non-real-time Services,'' \emph{KSII Trans Internet \& Information Systems}, Vol. 9, No. 8, pp. 2774--2796, Aug. 2015.

\bibitem{Dynamic2016Ji}
L. Chen, F. R. Yu, H. Ji, V. C. M. Leung, ``Dynamic Resource Allocation in Next Generation Cellular Networks with Full-Duplex Self-backhauls'', \emph{Wireless Networks}, 2016.

\bibitem{Resource2016Ji}
H. Ji, L. Xi, Z. He, W. Ke, `` Resource Allocation Scheme based on Game Theory in Heterogeneous Networks'', \emph{The Journal of China Universities of Posts and Telecommunications}, Vol. 23, No. 3, pp. 57--88, 2016.

\bibitem{Virtualization2018Yu}
F. R. Yu, J. Liu, Y. He, P. Si, and Y. Zhang, ``Virtualization for Distributed Ledger Technology (vDLT),'' \emph{IEEE Access}, 	Vol. 6, pp. 25019--25028, 2018.

\bibitem{5G2016Ge}
X. Ge, H. Wang, R. Zi, Q. Li and Q. Ni, ``5G Multimedia Massive MIMO Communications Systems,'' \emph{Wireless Communications and Mobile Computing (Wiley InterScience)}, Vol. 16, No. 11, pp. 1377--1388, Aug. 2016.

\bibitem{DelayA2017Ji}
Y. Zhi, W. Ke, H. Ji, ``Delay-aware Downlink Beamforming with Discrete Rate Adaptation for Green Cloud Radio Access Network'', \emph{The Journal of China Universities of Posts and Telecommunications}, Vol. 24, No. 1, pp. 26--34, 2017.


\bibitem{LabelLess}
Min Chen, V. Leung, ``From Cloud-based Communications to Cognition-based Communications: A Computing Perspective'', \emph{Computer Communications}, Vol. 128, pp. 74-79, 2018.

\bibitem{aiwac}
M. Chen, Y. Zhang, Y. Li, M. M. Hassan, A. Alamri, ``AIWAC: Affective Interaction Through Wearable Computing and Cloud Technology,'' \emph{IEEE Wireless Communications}, Vol. 22, No. 1, pp. 20-27, 2015.

\bibitem{Secrecy}
H. Long, W. Xiang, Y. Zhang, Y. Liu, W. Wang, ``Secrecy capacity enhancement with distributed precoding in multirelay wiretap systems,'' \emph{IEEE Transactions on Information Forensics and Security}, Vol. 8, No. 1, pp. 229-238, Jan. 2013.?

\bibitem{Privacy2017Guo}
H. Li, K. Wang, X. Liu, Y. Sun, S. Guo, ``A Selective Privacy Preserving Approach for Multimedia Data'', \emph{IEEE Multimedia Magazine}, Vol. 24, No. 4, pp. 14¡ª25, 2017.



\end{thebibliography}

\begin{IEEEbiography}[{\includegraphics[width=1in,height=1.25in,clip,keepaspectratio]{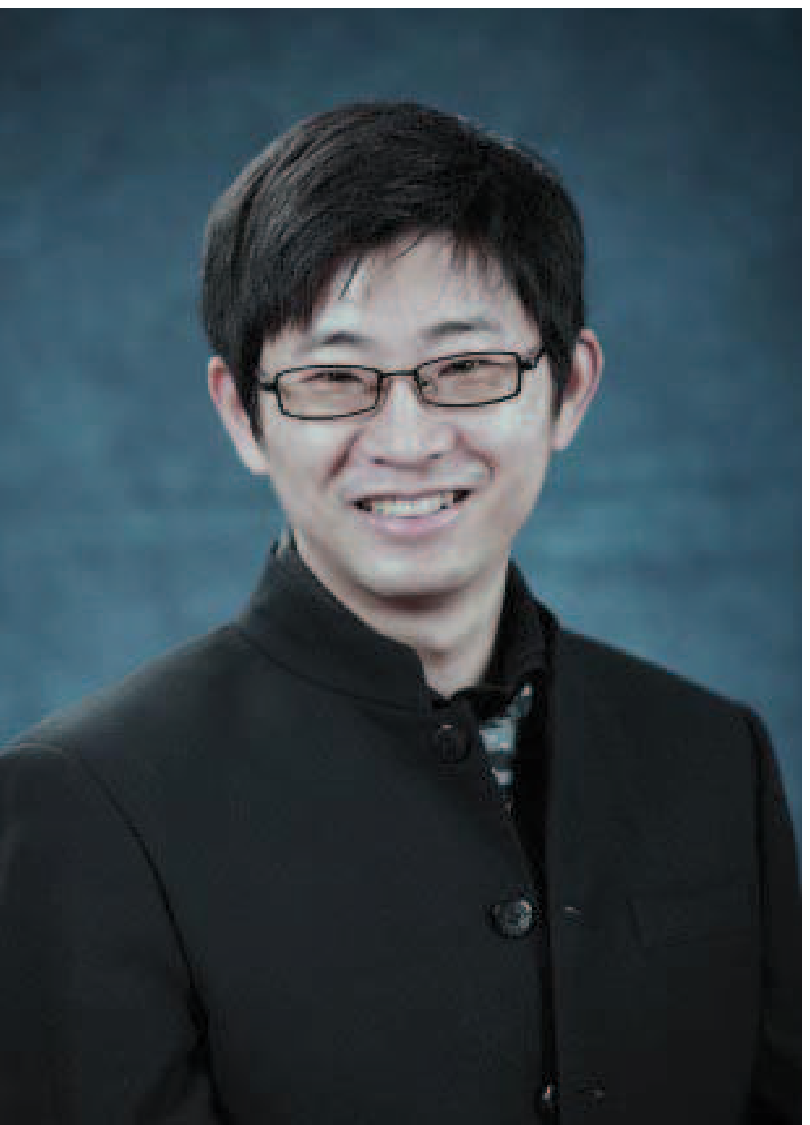}}] {Min Chen} [SM'09] has been a full professor in the School of Computer Science and Technology at HUST since February 2012. He is Chair of the IEEE Computer Society STC on big data. His Google
Scholars Citations reached 13500+ with an h-index of 58. He received the IEEE Communications Society Fred W. Ellersick Prize in 2017. His research focuses on cyber physical systems, IoT sensing, 5G networks, SDN, healthcare big data, etc.
\end{IEEEbiography}

\begin{IEEEbiography}[{\includegraphics[width=1in,height=1.25in,clip,keepaspectratio]{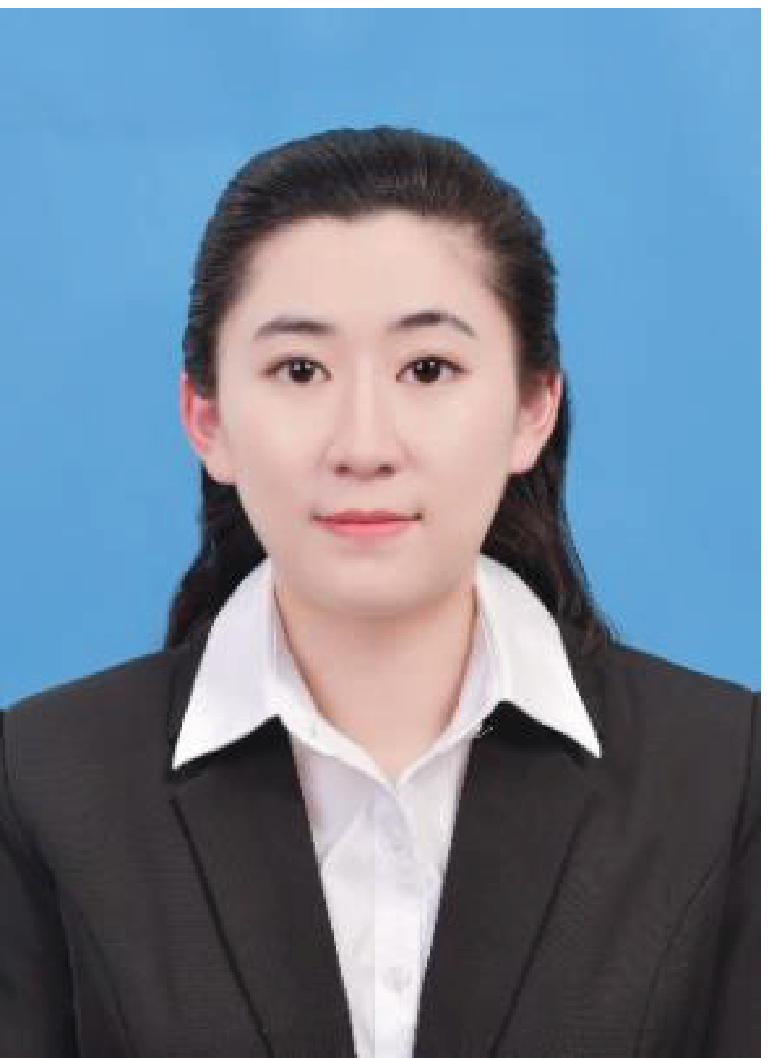}}] {Yiming Miao} received the B.Sc. degree in College of Computer Science and Technology from QingHai Univerisity, Xining, China in 2016. Currently, she is a Ph.D candidate in School of Computer Science and Technology at Huazhong University of Science and Technology (HUST). Her research interests include IoT sensing, healthcare big data and emotion-aware computing, etc.
\end{IEEEbiography}

\begin{IEEEbiography}[{\includegraphics[width=1in,height=1.25in,clip,keepaspectratio]{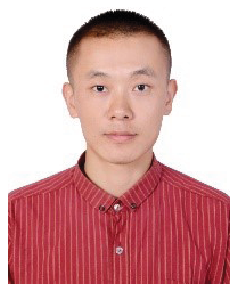}}] {Xin Jian} received his B.E. and Ph.D. degree from Chongqing University, Chongqing, China in 2009 and 2014, respectively. He is an associate professor at the College of Communication Engineering, Chongqing University, China. His interests include the next generation mobile communication, massive machine type communications, Narrow band Internet of Things.
\end{IEEEbiography}

\begin{IEEEbiography}[{\includegraphics[width=1in,height=1.25in,clip,keepaspectratio]{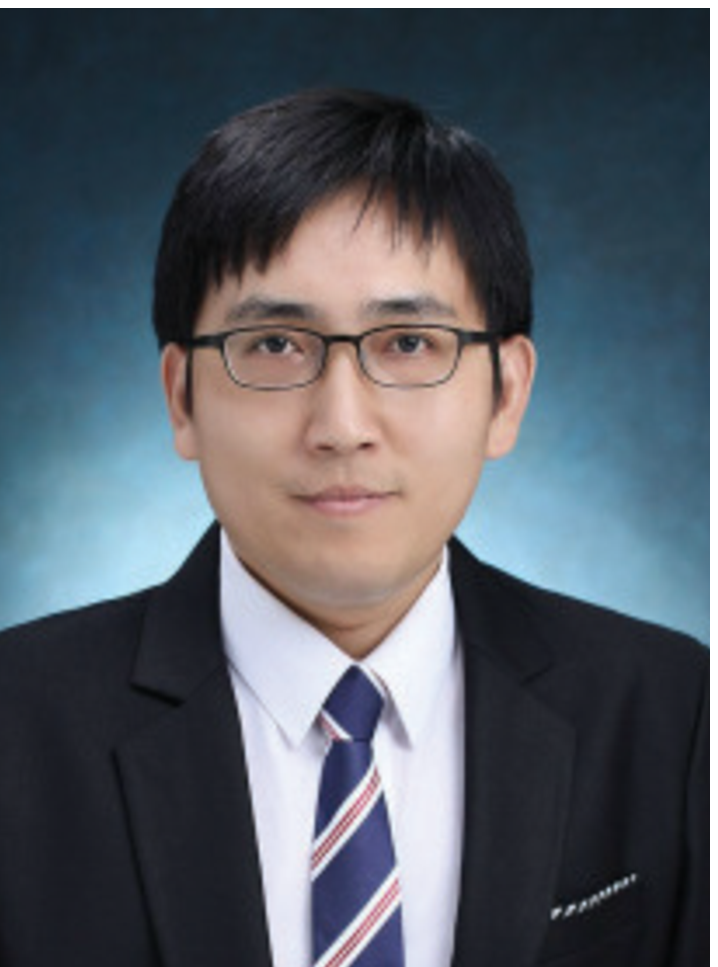}}] {Xiaofei Wang} [M'10, SM'18] received the MS and PhD degrees from the School of Computer Science and Engineering, Seoul National University, in 2008 and 2013, respectively. He worked as a post-doctoral research fellow in the University of British Columbia, Canada from 2014 to 2016. He is currently a professor with the School of Computer Science and Technology, Tianjin University, China. He has published more than 80 research papers in top journals and conferences, and got the IEEE Communications Society Fred W. Ellersick Prize, in 2017. His research interests include cooperative edge caching, and D2D traffic offloading. He is a senior member of the IEEE.
\end{IEEEbiography}

\begin{IEEEbiographynophoto} {Iztok Humar} [M'01, SM'10] received Ph.D. degrees in telecommunications from the Faculty of Electrical Engineering (FE) and information management at the Faculty of Economics, University of Ljubljana, Slovenia, in 2007 and 2009, respectively. He is an assistant professor at the FE, where he lecturers on design, management and modeling of telecommunication networks. His main research topics include the design, planning and management of telecommunications networks and services, as well as measurement and modeling of network loads and traffic. Currently, he is a visiting professor at Huazhong University of Science and Technology (HUST), China, where he works on energy-efficient wireless networks. Currently, he serves as IEEE Communication Society of Slovenia Chapter Chair, MMTC member, and IEEE Slovenia Section Secretary.
\end{IEEEbiographynophoto}

\end{document}